\documentstyle[twocolumn,aps,pra]{revtex}

\begin{document}

\wideabs{

\title{Dissipative dynamics of vortex arrays in trapped Bose-condensed gases: neutron
stars physics on \protect\( \mu K\protect \) scale.}

\author{P. O. Fedichev\protect\( ^{1,2}\protect \) and A. E. Muryshev\protect\( ^{2}\protect \)}

\address{\protect\( ^{1)}\protect \)Institut f. Theoretische Physik, Universitat Innsbruck,
Technikerstr. 25/2, A6020, Innsbruck, Austria}

\address{\protect\( ^{2)}\protect \)Russian Research Center ``Kurchatov Institute'',
Kurchatov Square, 123182, Moscow, Russia}

\maketitle
\begin{abstract}
We develop a theory of dissipative dynamics of
large vortex arrays in trapped Bose-condensed
gases. We show that in a static trap the
interaction of the vortex array with thermal
excitations leads to a non-exponential decay of
the vortex structure, and the characteristic
lifetime depends on the initial density of
vortices. Drawing an analogy with physics of
pulsar glitches, we propose an experiment which
employs the decay induced heating of the thermal
cloud as a tool for a non-destructive study of the
vortex dynamics. \end{abstract} }

Superfluidity is one of the striking
manifestations  of quantum mechanics in
interacting many body systems. The discovery of
Bose-Einstein condensation in dilute atomic gases
\cite{Cor95,Hul95,Ket95} provides a novel example
of superfluid systems. Until recently, the
evidence for superfluidity in these systems has
been fairly indirect and was based on either
excitation spectrum or measurements of critical
velocity (see
\cite{Ketterle:review,pitaevskii:review} for an
overview of extensive experimental and theoretical
studies). A  promising way of studying
superfluidity in trapped gases is the observation
of quantized  vortices
\cite{vortobserv:cornell,vortobserv:dalibard,vorto
bserv:foot}.

Aside from being a textbook example of persistent
currents, quantized vortices are believed to be
widespread in Nature. They play an important role
in breakup of superfluidity and superconductivity
\cite{LL:volIX}. More complicated structures, such
as arrays of vortex lines, appear in superfluids
rotating with a sufficiently large angular
velocity. This may occur in superfluid cores of
rotating neutron stars, where the dynamics of
vortices is thought to explain irregular features
in pulsars signals (so called glitches, see
\cite{alpar:glitchesI} for review).

Vortex arrays in trapped Bose-condensed  gases
have recently been observed at ENS
\cite{dalibard:varray}. Vortices were  formed in
the course of Bose condensation by stirring the
gas sample with laser light. Depending on the
angular frequency of the stirring beam, it was
possible to create configurations with different
number of vortices which form crystal-like
structures. The vortex array is visualized by a
destructive measurement of the gas cloud after
ballistic expansion. The experiment shows that
after the stirring ends, the vortex configuration
disintegrates in a clearly non-exponential
fashion.

The idea to model superfluid interiors of  neutron
stars in experiments with vortices in liquid
helium has a long history (see \cite{Donelly} for
an overview). Following this analogy, vortex
experiments in dilute gases may serve as an
``Earth-scale'' lab for studying the internal
dynamics of rotating neutron stars. Namely, the
Bose condensate filled with vortices can be used
as a model for the superfluid core of the star.
The thermal cloud interacting with the condensate
can play a role of the non-superfluid crust of the
star. We develop a theory explaining how the
interaction between the two components of the
system makes the vortices move towards the border
of the condensate, where they decay by creating a
sudden burst of thermal excitations (a
``glitch''). We propose an experiment which
employs the heating of the thermal cloud as a tool
for a non-destructive study of the vortex
dynamics.

For simplicity, we consider a large array of
vortex lines, emerged in a long cylindrical 
magnetic trap characterized by the radial trap
frequency \( \omega  \). Further on we assume that
the angular velocity of the trap rotation \(
\Omega \gg \Omega _{c} \), where \( \Omega _{c} \)
is the critical velocity corresponding to the
appearance of the first vortex. In this case the
equilibrium number of vortices \( N\gg 1 \). We
assume that the stirring beam is on for a
sufficiently long time so that the system reaches
thermal equilibrium in the rotating frame. The
equilibrium configuration of the system in the
rotating frame can be obtained by variational
analysis of the free energy functional
\begin{equation} \label{freeen} {\mathcal{F}}=\int
d^{3}rn_{0}({\textbf {r}})\left( \frac{m{\textbf
{v}}_{s}^{2}}{2}+V({\textbf {r}})-\mu
+\frac{g}{2}n_{0}({\textbf {r}}))-(\Omega
,{\textbf {M}})\right) , \end{equation}  where \(
V({\textbf {r}})=m\omega ^{2}r^{2}/2 \) is the
trapping potential in the radial direction, \(
g=4\pi \hbar ^{2}a/m \), \( m \) is the the atom
mass, \( \mu  \) is the chemical potential, and \(
n_{0}({\textbf {r}}) \), \( {\textbf {v}}_{s} \)
and $$  {\textbf{M}}=\int dr^3 mn_0 {\bf
r}\times{\bf v}_s $$ are, respectively, the
superfluid density, velocity and the angular
momentum of the condensate in the
\textit{laboratory} reference frame. The
expression for the free energy given by
Eq.(\ref{freeen}) implies that the condensate is
sufficiently dense: the mean interpaticle
interaction \( \mu  \) is much larger than the
level spacing in the trap \( \hbar \omega  \)
(Thomas-Fermi limit).

In the presence of a vortex array with the vortex 
density \( n \), the superfluid velocity satisfies
Maxwell-like equations \cite{LL:volIX}
\begin{equation} \label{Maxwell}
{\rm rot}{\textbf {v}}_{s}=2\pi \kappa n\hat{z},\;
{\rm div}{\textbf {v}}_{s}=0, \end{equation}
 where \( \hat{z} \) is a unit vector along  the
axis of the cylinder , and where \( \kappa =\hbar
/m \) is the vortex circulation. The superfluid
velocity inside a cylindrically symmetric vortex
array can be found from Eq.(\ref{Maxwell}) by
using the Stokes theorem. It has no radial
component (i.e. it satisfies the necessary
boundary condition at the border of the
condensate) and is given by \begin{equation}
\label{vs} {\bf v}_{s}({\bf r})=2\pi \frac{[\kappa
\times {\bf \hat{r}}]}{r}\int _{0}^{r}r^{\prime
}dr^{\prime }n(t,r^{\prime }), \end{equation}
where \( {\bf {\hat{r}}} \) is the unit vector in
the direction of \( {\textbf {r}} \).

The equilibrium configuration of \( n_{0}({\textbf
{r}}) \) and \( {\textbf {v}}_{s} \) minimizes the
free energy \( {\mathcal{F}} \). The variational
solution has to be found within the class of
functions \( {\textbf {v}}_{s} \) corresponding to
a given total number of vortices. Variation of the
free energy (\ref{freeen}) with respect to the
vortex velocity shows that in thermal equilibrium 
the velocity of the superfluid flow imitates the
flow of a normal liquid with the angular velocity
\( \Omega  \): \( {\textbf {v}}_{s}=[\Omega \times
{\textbf {r}}] \). The corresponding density of
vortex lines is also spatially uniform and is
given by \cite{LL:volIX} \begin{equation}
\label{eqdens} n=\frac{\Omega }{\pi \kappa }.
\end{equation}  Then, the minimum of the free
energy is reached at the condensate density given
by the following modified Thomas-Fermi expression:
\[ n_{0}({\textbf {r}})=n_{0}(0)\left(
1-\frac{r^{2}}{R_{c}^{2}}\right) ,\]  where the
transverse size of the condensate is given by \[
R^{2}_{c}=\frac{2\mu }{m(\omega ^{2}-\Omega
^{2})},\] and \( \mu =gn_{0}(0) \). We note that
the angular velocity \( \Omega  \) can not be very
large: The maximum velocity of the superflow,
which coincides with the velocity of the
superfluid on the border of the condensate is \(
\sim \Omega R_{c} \) and should not exceed the
velocity of Bogolyubov sound \( c_{s}\sim
\sqrt{\mu /m} \). We thus have
\begin{equation} \label{omegamax} \Omega \ll
\frac{c_{s}}{R_{c}}\sim \omega , \end{equation}
since otherwise the superflow becomes unstable
with respect to spontaneous creation of
excitations (see Landau arguments in
\cite{LL:volIX}). This limits the maximum number
of vortices inside the vortex array: \( N\alt
N_{max}=\mu /\hbar \omega  \), which is still a
large number in the Thomas-Fermi limit. In what
follows we will assume that the number of vortices
is still sufficiently small \( 1\ll N\ll N_{max}
\), so that the angular velocity of the
condensate rotation is smal compared to the
trap frequency. Therefore, we can neglect the
size increase of of the condensate due to the
rotation.

When the rotation of the trap ceases (i.e. after
the stirring light is switched off and the
rotation of the thermal cloud stops, see
\cite{godelin:kinetics}), the vortex array
continues to rotate with the angular velocity \(
\Omega  \). The vortex configuration is described
by two continuous functions: the average
vortex density \( n(t,{\textbf {r}}) \) and
velocity \( {\textbf {v}}_{v}(t,{\textbf {r}}) \).
Since the number of vortex lines is a locally
conserved quantity, the introduced functions
satisfy the continuity equation: \begin{equation}
\label{conteq} \frac{\partial n}{\partial
t}+\frac{\partial ({\textbf {v}}_{v}n)}{\partial
{\textbf {r}}}=0. \end{equation}  With this
``macroscopic'' description of the vortex
``matter'', we take a qualitative view on the
vortex dynamics, ignoring the issues of metastable
vortex configurations and the nature of elementary
excitations in the vortex array (see
\cite{Donelly} for a more detailed discussion).

The vortex velocity is related to the superfluid
velocity through the Magnus law \cite{Sonin}:
\begin{equation}
\label{MagnusLaw}
\rho _{s}[{\textbf {v}}_{v}-{\textbf {v}}_{s},
\kappa ]={\textbf {F}}, \end{equation}
 where \( \rho _{s} \) is the superfluid (mass)
density and \( {\textbf {F}} \) is the friction
force acting on the vortex line (per unit length).
At finite temperatures, the force \( {\textbf {F}}
\) originates from the scattering of thermal
excitations from moving vortices and consists of
two terms: \begin{equation} \label{frictionforce}
{\textbf {F}}=-D{\textbf {v}}_{v}-D^{\prime
}[\kappa \hat{z},{\textbf {v}}_{v}].
\end{equation}  Here \( D \) and \( D^{\prime } \)
are the longitudinal and transverse friction
coefficients, respectively. Both of them are
temperature dependent and \( D^{\prime }=\rho _{n}
\), where \( \rho _{n} \) is the density of the
normal component. The discussion of the friction
coefficient \( D \) can be found in
\cite{Sonin,fedichev:vortexdyn}. The solution of
Eq.(\ref{MagnusLaw}), with \( {\textbf {v}}_{s} \)
from Eq.(\ref{vs}), can be represented in the
form: \( {\bf {v}}_{v}=v^{(r)}{\bf
{\hat{r}}}+v^{(\phi )}[\hat{\kappa }\times {\bf
{\hat{r}}}] \), where \begin{equation} \label{vv}
v_{v}^{(r)}=v_{s}\frac{\kappa \rho _{s}D}{\kappa
^{2}\rho ^{2}+D^{2}},\; v_{v}^{(\phi
)}=v_{s}\frac{\kappa ^{2}\rho \rho _{s}}{\kappa
^{2}\rho ^{2}+D^{2}} \end{equation}  are the
radial and tangential components of the velocity
field, and \( \rho =\rho _{s}+\rho _{n} \) is the
total density. Typically \( D\ll \kappa \rho  \)
and \( \rho \approx \rho _{s} \) (see below).
Hence, we have \( v_{v}^{(r)}\approx Dv_{s}/\kappa
\rho \ll v_{s} \) and \( v_{v}^{(\phi )}\approx
v_{s} \). This means that in the presence of a
finite friction (\( D\neq 0 \)) the vortices
circulate around the trap center with velocity \(
v_{s} \) and slowly spiral out towards the walls
of the vessel. At the border of the trap the
condensate density vanishes and the vortex decays,
creating a burst of excitations and thus
transferring its energy to the thermal cloud.

Assuming that the vortex density does not depend
on the polar angle, i.e. \( n(t,{\textbf
{r}})=n(t,r) \), and using Eqs.(\ref{vs}) and
(\ref{vv}), we rewrite Eq.(\ref{conteq}) in the
form

\begin{equation}
\label{disdynmodel}
\frac{\partial n(t,r)}{\partial t}+\frac{\pi
\kappa  \alpha }{r}\frac{\partial }{\partial
r}\left( n(t,r)\int _{0}^{r}r^{\prime }dr^{\prime
}n(t,r^{\prime })\right) =0, \end{equation}  where
\( \alpha =2\kappa \rho _{s}D/(\kappa ^{2}\rho
^{2}+D^{2})\approx 2D/\kappa \rho  \). Although
this equation is not local, the density at the
point \( {\textbf {r}} \) depends only on the
density profile inside the cylinder of radius \( r
\).

The solution of Eq.(\ref{disdynmodel}), with the
initial density profile given by
Eq.(\ref{eqdens}), remains spatially uniform at
any \( t \) (\( n(t,r)=n(t) \)) and satisfies the
equation \[ \frac{dn(t)}{dt}+\pi \kappa \alpha
n^{2}(t)=0.\]  Given the initial condition \(
n(0)=n_{eq} \), the total number of vortices \(
N(t)=\pi R^{2}n(t) \) decays non-exponentially:
\begin{equation} \label{Nintime}
N(t)=\frac{N(0)}{1+t/\tau _{R}}.
\end{equation}
 Here \( N(0)=\pi R^{2}n_{eq} \), and the
characteristic  relaxation time \( \tau _{R} \) of
the vortex structure depends on the initial
density \begin{equation} \label{taurel}
\tau _{R}^{-1}=\pi \kappa \alpha n_{eq}=\pi \kappa
n_{eq}\frac{2\kappa \rho _{s}D}{\kappa ^{2}\rho
^{2}+D^{2}}. \end{equation}

This result makes sense even if the number of
vortices  is not large. For example, in the case
of a single vortex, i.e. for \( N=1 \), we have \(
n_{eq}\sim 1/R^{2} \), where \( R \) is the size
of the vessel. Then, by substituting this density
into Eq.(\ref{taurel}) we find that \( \tau
_{R}^{-1}\sim \pi \alpha \hbar /mR^{2} \). This
differs only by a logarithmic factor from the
life-time of a single vortex in a static trap
\cite{fedichev:vortexdyn}. The similarity of the
results is not surprising, since the scenarios of
dissipative dynamics in both cases are
qualitatively the same.

To be more specific, we present explicit
expressions  for \( \tau _{R} \) in a weakly
interacting Bose-condensed gas. First we consider
the limit of very small temperatures \( T\ll \mu
\), where \( \mu =4\pi \hbar ^{2}an_{0}/m \) is
the chemical potential of the gas, \( n_{0} \) is
the condensate density, and \( a \) is the 2-body
scattering length. Then, since \( \rho _{S}\approx
mn_{0} \), and \( D\propto \kappa m^{5/2}T^{5}/\mu
^{7/2}\hbar ^{3} \) \cite{Donelly}, we find that
\( D\ll \kappa \rho  \) and \begin{equation}
\label{taurlowT} \tau _{R}^{-1}\propto
\frac{\kappa N(0)}{\pi R^{2}}\left(
n_{0}a^{3}\right) ^{1/2}\left( \frac{T}{\mu
}\right) ^{5}. \end{equation}

In the opposite limit (\( \mu \ll T \)) the
calculation of the friction coefficient gives \(
D\approx 0.1\kappa m^{5/2}T\mu ^{1/2}/\hbar ^{3}
\) \cite{fedichev:vortexdyn}. Then, assuming \(
\rho _{n}\ll \rho  \), we obtain \begin{equation}
\label{taurhigh} \tau _{R}^{-1}\approx 1.6\pi
^{5/2}\frac{\kappa N(0)}{\pi
R^{2}}(n_{0}a^{3})^{1/2}\frac{T}{\mu }.
\end{equation}  In a spatially homogeneous gas the
condition \( \rho _{n}\ll \rho  \) used to derive
Eq.(\ref{taurhigh})implies \( T\ll T_{c} \), where
\( T_{c} \) is the temperature of Bose-Einstein
condensation. In a harmonic trap characterized by
the frequency \( \omega  \), thermal particles and
the condensate occupy different volumes: \(
V_{T}=(2T/m\omega ^{2})^{3/2} \) and \(
V_{c}=(2\mu /m\omega ^{2})^{3/2}\ll V_{T} \),
respectively. Hence, the density of the thermal
component can be much smaller than \( n_{0} \)
even at temperatures close to \( T_{c} \). This
condition requires the fraction of Bose-condensed
particles, which is \( \sim (T-T_{c})/T_{c} \), to
be larger than \( V_{c}/V_{T}=(\mu
/T_{c})^{3/2}\ll 1 \) (see
\cite{pitaevskii:review}).

Given the values of experimental parameters
\cite{dalibard:varray}: \( \omega =169\times 2\pi
s^{-1} \) and the number of particles \( 10^{5},
\) we find that \( \mu \approx 100nK \) and \(
R=(2\mu /m\omega ^{2})^{1/2}\approx 5\mu m \). At
temperatures \( T\sim \mu  \) the size of the
thermal cloud matches the size of the condensate.
This prevents further evaporative cooling and
complicates the temperature measurements. For this
reason, we estimate the relaxation time for \(
T=\mu  \), where both Eqs. (\ref{taurlowT}) and
(\ref{taurhigh}) give approximately the same
result: \[ \tau _{R}\sim 4/N(0)\quad s.\]  For the
array of \( 6 \) vortices we estimate \( \tau
_{R}\approx 0.7s \), which is in good agreement
with the measured value \( \approx 1s \).

Vortices in trapped Bose-condensed gases are very
small and can not be visualized non-destructively
by optical means. This is similar to the situation
in astrophysics, where indirect evidence for the
vortex dynamics was obtained by observing
irregular features in signals of pulsars
associated with rotating neutron starts
\cite{alpar:glitchesI}. Interiors of neutron stars
may contain superfluid nuclear matter sustaining
huge vortex arrays. Every time a vortex reaches
the crust of the star, it decays and the sudden
release of its energy is detected on Earth as a
glitch in an otherwise periodic signal of the
star.

In trapped gases the vortex decay transfers vortex
energy to the thermal cloud and hence increases
the gas temperature. As we have noted, the
superfluid flow with a large number of vortices
imitates the rotation of a normal liquid.
Therefore the energy of a vortex array can be
considered as an energy of rotation of such a
liquid: \( E=I\Omega ^{2}/2 \), where \( I=\pi
mn_{0}R_{z}R^{4}/2 \) is the momentum of inertia
of a condensate, \( R_{z}=\sqrt{2\mu /m\omega
_{z}} \) and \( \omega _{z} \) are the condensate
size in axial direction and the axial trap
frequency, correspondingly. Given the heat
capacity of the gas cloud \( C\propto T^{3}/\hbar
^{3}\omega ^{2}\omega _{z} \), we find that the
temperature increase after the decay the vortex
array: \[ \frac{\Delta T}{T}\approx
\frac{E}{CT}\sim
\frac{1}{(n_{0}a^{3})^{1/2}}\left( \frac{\mu
}{T}\right) ^{4}\left( \frac{\Omega }{\omega
}\right) ^{2}.\]
The quantity $\Delta T$  can
be large, at least for the temperatures well
below the Bose-Einstein condensation temperature
\( T_{c} \). The heating of a Bose-condensed gas,
associated with the decay of the vortices, can be
detected by a non-destructive technique.  Then,
for instance, by plotting the temperature of the
thermal cloud as a function of time, one can
confirm the decay scenario described by
Eq.(\ref{Nintime}). This may provide a new tool
for the study of the vortex dynamics and, at the
same time, serve as a laboratory model of a
rotating neutron star.

In conclusion, we have developed a theory of
dissipative dynamics of large vortex arrays. The
calculated characteristic lifetime of the vortex
structure in a static trap depends on the initial
density of vortices and agrees well with the
recent experiment \cite{dalibard:varray}. Putting
forward the analogy between the dissipative
dynamics of vortices in trapped gases and the
physics of neutron stars, we propose an
experiment, where vortices can be indirectly and
non-destructively studied by measuring  the
heating of the thermal component

We acknowledge fruitful discussions with G.V.
Shlyapnikov, J. Dalibard, P. Zoller, J.I. Cirac,
and I.E. Shvarchuck. The work was supported by
Austrian Science Foundation, by INTAS, and Russian
Foundation for Basic Research (grant 99-02-18024).
\bibliographystyle{prsty}
\bibliography{/home/fedichev/LyX/myDb}

\end{document}